# Flight demonstration of formation flying capabilities for future missions(NEAT Pathfinder)

M. Delpech [1], F.Malbet [2], T.Karlsson [3], R.Larsson [3], A.Léger [4], J.Jorgensen [5]

(1) CNES, 18 avenue Edouard Belin, 31041 Toulouse, France
(2) IPAG, 414 rue de la Piscine, 38400 St Martin d'Hères, France
(3) OHB Sweden, Solna Strandvaeg 86, 17122 Solna, Sweden,
(4) IAS, University Paris Sud, Orsay, France
(5) DTU Space, Elektrovej, Building 327, DK-2800 Lyngby, Denmark,

**ABSTRACT**

PRISMA is a demonstration mission for formation-flying and on-orbit-servicing critical technologies that involves two spacecraft launched in low Earth orbit in June 2010 and still in operation. Funded by the Swedish National Space Board, PRISMA mission has been developed by OHB Sweden with important contributions from the German Aerospace Centre (DLR/GSOC), the French Space Agency (CNES), and the Technical University of Denmark (DTU). The paper focuses on the last CNES experiment achieved in September 2012 that was devoted to the preparation of future astrometry missions illustrated by the NEAT and µ-NEAT mission concepts. The experiment consisted in performing the type of formation maneuvers required to point the two-satellite axis to a celestial target and maintain it fixed during the observation period. Achieving inertial pointing for a LEO formation represented a new challenge given the numerous constraints from propellant usage to star tracker blinding. The paper presents the experiment objectives in relation with the NEAT/µ-NEAT mission concept, describes its main design features along with the guidance and control algorithms evolutions and discusses the results in terms of performances achieved during the two rehearsals.

# INTRODUCTION

Formation flying (hereafter FF) does constitute the leading approach to build large dimensions instruments while getting free from most launchers payload accommodation constraints. Numerous mission concepts have been proposed to satisfy a large diversity of objectives from the Sun observation by a bi-satellite coronograph (Proba3 [1]) up to the detection and observation of exoplanets with 4 or 5 satellite interferometers (Darwin [2]). These scientific missions that imply inter spacecraft separations from tens to hundreds of meters have common features: they impose stringent constraints on the relative positioning / pointing of the vehicles as well as their stabilization and this require the accommodation of new fine metrology and actuation systems.

Despite the high interest of the worldwide space community for these emerging technologies and the exciting perspectives offered for the scientific exploration of the Universe, formation flying suffers from a severe prejudice: its implementation appears too expensive and far too complicated. Hence, soon after some apex at the beginning of the century when ambitious mission concepts were flourishing, formation flying started to vanish from the programmatic landscape and projects already initiated in Europe such as Simbol-X [3], XEUS and also Darwin were actually interrupted (NASA Starlight mission [4] had already known that fate a few years earlier). Nowa-days, platform design is bending to a more conservative side with the systematic use of deployable masts whenever large dimension instruments have to be developed. Of course, these short-term technical solutions do have significant benefits since off-the-shelf products already exist and allow sticking within the single spacecraft development paradigm, which is easier to manage from the project organisation point of view. In this context, formation flying complexity seems overestimated and the barrier could be somewhat psychological. Numerous innovative missions have been previously decided even though critical components relied only on ground proven technology. Conversely, FF mission concepts are expected to remain in the waiting room, as long as the whole set of required equipment and functionalities have not been flight proven. This seems to be the fate of some new astrometry missions like NEAT [5] and its precursor µ-NEAT despite their excellent scientific return - complexity ratio. Significant preparatory steps have been accomplished, though, by PRISMA technology mission [6] in Low Earth orbit that allowed demonstrating formation flying technological components and representative operations. In particular, formation flying operations were performed in routine manner with RF metrology (that is the candidate metrology stage for outer space) along with the relevant guidance, navigation and control functionalities. The achieved level of performance was obviously below the requirements of future mission concepts since the equipment suite did not involve any fine actuation or adequate optical metrology and the gravity gradient environment was far from favourable. In addition, the impact might have been limited because the demonstration was too closely related to satellite servicing activities. Therefore, ESA Proba3 mission that aims at demonstrating fine positioning on some High Elliptical Orbit appears to be a valid opportunity to break the conservatism and opens an era of more ambitious missions.

The work presented in this paper represents a modest effort to bridge the gap between the demonstrations achieved so far with PRISMA in the field of formation flying astronomy and the accomplishments expected from a future technology mission like Proba3. The work consists in demonstrating in flight with PRISMA platforms a scenario of formation manoeuvres that would be performed in the NEAT / µ-NEAT mission concepts. Given the constraints of the PRISMA system and the orbital environment, the ambition is not to reproduce accurately the configuration and try to satisfy the positioning and pointing requirements. The purpose is to focus on the functional aspects while pushing the PRISMA system to its operational limit. To enhance the illustrative impact, the experiment involves the pointing of the formation to a selection of scientific targets belonging to the actual NEAT/µ-NEAT catalogue. The experiment has been actually nicknamed "NEAT pathfinder" to outline its preparation objective.

The paper presents in the next section the main characteristics of the NEAT/µ-NEAT mission concepts. Then, the PRISMA system and its current capabilities are described in a third section. The detailed design of the experiment is developed in the fourth section with a discussion on its current limitations and constraints and a special focus on the guidance and control issues. The presentation of the flight results including a discussion on the navigation and control performances along with camera images illustrating the inertial pointing is given in the fifth section. Finally, the conclusion summarizes the different experiment achievements and opens perspectives for future work.

**NEAT / µ-NEAT MISSION CONCEPTS**

NEAT / µ-NEAT are two astrometric missions based on formation flying that have been proposed to ESA with different scales: NEAT as a M-class mission with a 1 m telescope an µ-NEAT as a S-class mission with a 0.3 m telescope. The experiment has been proposed to prepare and illustrate the NEAT / μ-NEAT mission concepts in the high-precision astrometry field dedicated to the detection and characterization of planets in the so-called "Habitable zone" (0~.5-2 AU around solar-type stars). NEAT is focused on Earth-like planets whereas μ-NEAT is aimed at larger planets (from 1 to 50 times the Earth size). The NEAT/μ-NEAT concepts are based on a simple optics with only one off-axis parabolic mirror (called the telescope) and a focal plane located at the distance that allows correctly sampling the point spread function. The differential astrometry measurement is performed by an external metrology system that can also calibrate precisely the detector pixel responses. The objective is to reach micro-pixel accuracy. The main characteristics of NEAT / μ-NEAT are presented here below:

- L2 Lagrange point mission: 5 years (NEAT) / 3 years (μ-NEAT)
- Formation flying with 2 spacecraft (Telescope and Detector)
- Spacecraft distance: 40 m (NEAT) / 12 m (μ-NEAT)
- Up to 20 reorientation manoeuvres per day
- Telescope pointing accuracy: 3 arcsec
- Detector positioning accuracy: 2 mm on all axes (NEAT) / 1 cm (μ-NEAT)
- Observation lasts between 30min and a few hours
- 200 targets visited 50 times
- Average angle distance between 2 pointing directions: 10°

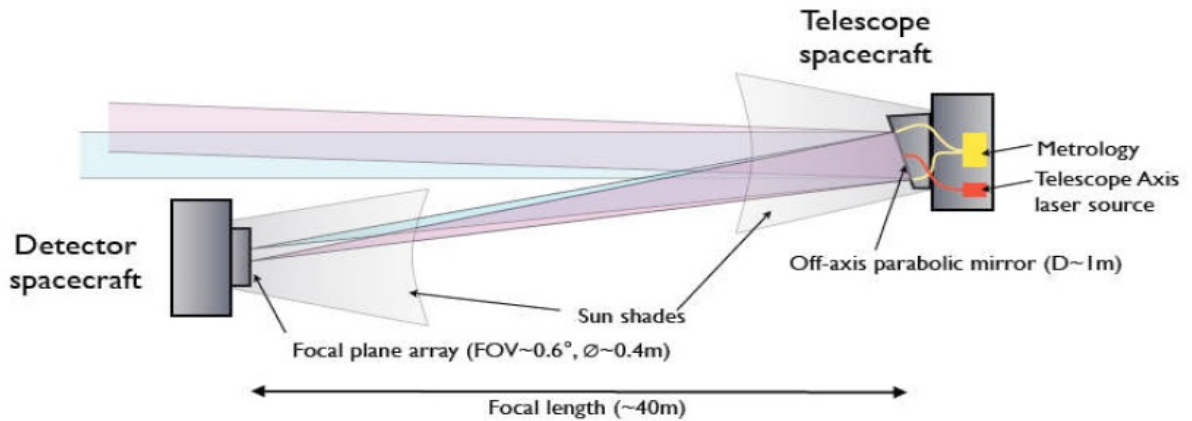

*Figure 1: schematics of the NEAT spacecraft concept*

**PRISMA SYSTEM DESCRIPTION**

The PRISMA space segment launched in June 2010 on a 700 km sun-synchronous orbit consists of a small satellite Mango (150 kg), and a microsatellite Tango (40 kg). Mango has full 3-dimensional attitude independent orbit control capability and is 3-axis attitude stabilized using star trackers and reaction wheels. Tango does not have any attitude control capability and is equipped with a solar magnetic attitude control system still providing 3- axis stabilization. The propulsion system on Mango is based on six 1-N thrusters directed through the spacecraft centre of mass and the delta-V capability is approximately 120 m/s. All ground communication is made with Mango and communication with Tango is made via an inter-satellite link (ISL). The flight system is developed by OHB-Sweden but includes several hardware / software contributions from European partners. DLR provided GPS receivers on both satellites, a relative navigation system and a dedicated experiment software to perform autonomous formation keeping [7]. DTU delivered a set of Vision Based Sensors (VBS) that enabled optical navigation in non-cooperative mode (far range VBS) and cooperative mode (close range VBS) [8]. CNES provided a new Formation Flying Radio Frequency (FFRF) metrology sub-system designed for future outer space FF missions and some on-board GNC software to perform a variety of formation flying experiments [9]. As a service to all experimenters, DLR provided also Precise Orbit Determination (POD) that is computed on the ground from raw GPS measurements and serve as positioning reference given its high accuracy.

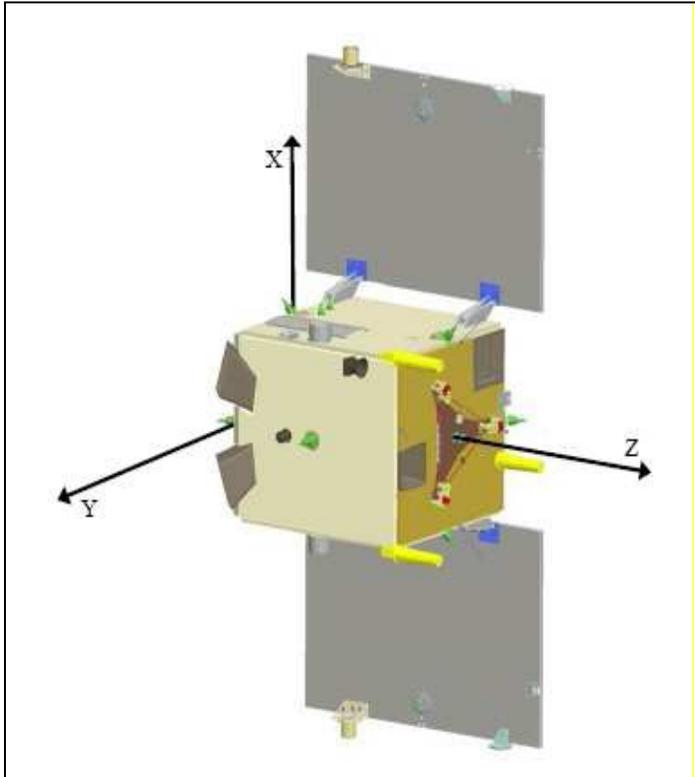
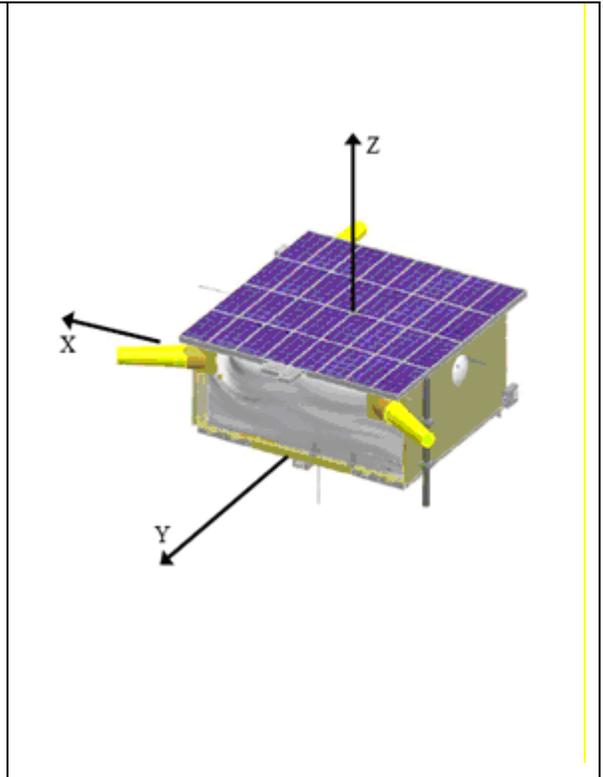

**Figure 2.1: Mango satellite**. Cameras and FFRF antennas (yellow cylinders are located on the spacecraft Z face.

**Figure 2.2: Tango satellite**

Mango carries 3 cameras looking along the Z axis: 2 black and white VBS cameras that can be used in closed loop and a colour camera (DVS) accommodated for public outreach purpose. Multiple options are therefore available to illustrate the relative configuration as well as the inertial pointing.

Thales Alenia Space has developed the FFRF Sensor that is used in this experiment with a partnership from CDTI. It is a distributed instrument designed to provide range and Line Of Sight (LOS) measurements at 1 Hz with a typical accuracy of 1 cm and 1°. Its functional principle is inherited from the Topstar GPS using dual frequency S-band signals [10]. Requirements included also the lost in space capability assuming some indirect assistance from the GNC system for signal ambiguity resolution (IAR). Besides, its generic initial design was adapted to fit the accommodation constraints on PRISMA (10 kg and 30 W) and to take into account the formation asymmetric structure composed of a smart chaser satellite (Mango) and a passive target (Tango). In addition, the instrument offers an inter-satellite link (ISL). This capability, not part of the nominal PRISMA communication system, was exercised only for validation.

Numerous proximity operations were performed using GPS, VBS and FFRF sensors [11]. These allowed to determine the level of positioning accuracy achievable in station keeping. 3D control stability close to 1 cm (1σ) was demonstrated at 20 m range with FFRF and using a particular actuation mode (sub-pulse mode) where thrusters are fired in a differential way to reduce the Minimum Impulse Bit (MIB) [12].

Similar performance was also obtained with VBS at shorter range and this was regarded as the ultimate achievable accuracy. It must be outlined that this level of performance was obtained with a constant formation pointing in LVLH frame whereas no experiment was performed so far in inertial pointing mode.

**DESCRIPTION OF THE EXPERIMENT**

**Formation configuration**

The set-up is tailored to approach the NEAT configuration but with some adaptation since the PRISMA position controlled satellite will play the role of Telescope spacecraft whereas the passive one will be the Detector. In addition, PRISMA satellites are not capable to achieve the NEAT/µ-NEAT mission performance requirements given the propulsion system resolution (MIB = 0.7 mm/s) and the large gravity gradient of the LEO environment. This leads to scale down the fidelity in several areas:

(1) *Accurate telescope pointing (3 arcsec):* this capability is not formation flying specific and has already been demonstrated in other missions in L2 like Herschel.

(2) *relative positioning accuracy (2 mm)*: This requires some precise optical metrology and fine resolution thrusters (ex: cold gas) not available on PRISMA. Despite the LEO context, an accuracy close to 1 cm 3D has been already achieved on PRISMA and this represents a fair indicator that such a requirement will be properly satisfied during the NEAT mission with appropriate sensors and actuators. However, demonstrating the station-keeping functionality in inertial configuration is achievable in LEO with a coarser accuracy and represents an additional challenge since this was never performed during the PRISMA mission.

(3) *Re-pointing manoeuvres*: The typical task that consists in switching from one target to the next is performed in several steps that must be coordinated: (#1) rotating the Telescope to the new target, (#2) translation of the Detector to position it along the new reflected beam at 40 m, (#3) rotating the Detector to face the Telescope, (#4) acquisition of the payload metrology and switch to fine navigation In the real mission, the synchronization of both spacecraft rotations is not mandatory and it is possible to operate each spacecraft independently: the Telescope performs its rotation (step #1) while the Detector moves to its destination (step #2) and rallies the desired attitude (step #3). Step #4 is then performed when both spacecraft have completed their manoeuvres. Steps #2 and #4 only represent a real formation flying activity. Step #2 requires relative navigation (radio-frequency metrology) and the ability to follow a desired relative trajectory in the inertial frame.

Here, the NEAT Pathfinder experiment focuses on the Detector spacecraft manoeuvres and particularly the translational part (step #2) plus the observation period. The intent is to be representative in terms of manoeuvre geometry and duration as well. The relative distance is 12 m (µ-NEAT configuration) and final accuracy target is in the 10 cm range during the observation phases. Manoeuvre success will be evaluated using on-board attitude and Precise Orbit Determination (POD) data. In addition, the manoeuvre completion is be illustrated by VBS images of the Detector spacecraft (Tango) and images of the sky centred on the NEAT target.

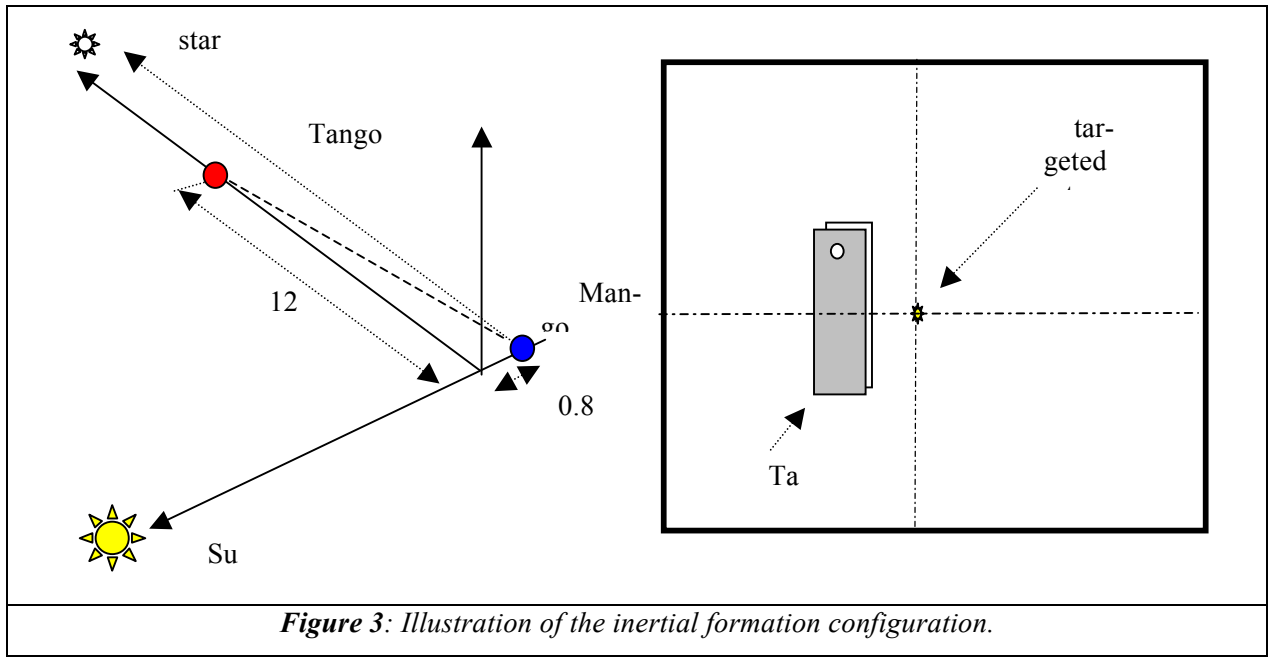

*Figure 3: Illustration of the inertial formation configuration.*

According to the NEAT configuration, Mango is not along the line defined by Tango and the targeted star. In the experiment context, it is located in the plane defined by the Sun, Tango, star triplet and the apparent angle between Tango and the targeted star is 3.8° (0.8 m transversal bias at 12 m). The left part of the figure represents the associated view as seen by Mango camera (VBS). Tango nominal pointing imposes the solar panel to be parallel to the orbital plane but variations in the 10-15° are expected.

**Limitations and constraints**

The main constraint in terms of attitude configuration and control comes from the following requirements: (1) perform relative control using as much as possible radio frequency metrology (FFRF) since this will be the first metrology stage in future missions; (2) achieve the best possible control performances in the various pointing configurations, (3) point to a selection of NEAT scientific targets that are potentially visible with on board instruments.

FFRF metrology highest accuracy is actually obtained when Mango S/C antennas can be aligned with one of the Tango S/C Rx/Tx antennas since this allows to reduce RF signal bias due to multipath. Some misalignment must be tolerated though given Tango S/C coarse attitude performances (about 10°). However, it is compulsory to maintain the misalignment low enough to avoid any antenna switch during the experiment (+/-60°). The ideal approach that was envisioned consisted in pointing both Mango S/C and Tango S/C in a coordinated way when aiming at a particular celestial target: a specific antenna would be selected and its axis would become the pointing reference direction for the whole experiment. In addition, this would make the overall attitude configuration more representative of the targeted missions. However, Tango S/C attitude being controlled by magnetic actuators, analysis outlined the difficulty to achieve sufficient stability margins in any pointing configuration while satisfying reasonable control performances. It was therefore decided to avoid the risk of attitude failure and stick to a less ambitious approach: Tango S/C is maintained in the same inertial attitude throughout the whole experiment and the list of candidate celestial targets is reduced in such a way that the relative pointing constraints can be satisfied. Figure 4 shows Tango S/C attitude configuration that offers a satisfactory stability in inertial pointing (Tango X axis is tilted 9° w.r.t the projection of ECI frame Z axis in the orbital plane).

Figure 4.2 shows a map of the sky including stars with magnitude < 6 when Mango S/C is located at the ascending node. The blue line represents the Earth limb whereas the red line is the limit of the Sun zone assuming a 30° exclusion angle. The sky zones that are observable when Tango satellite is pointed inertial (9° / Zenith) and when Mango is maintained in a 30° half angle cone around the antenna axes to satisfy the FFRF navigation requirement are represented by the three ellipses (one for each antenna). The most populated zone corresponds to the antenna FFRF_MY and this is the reason why it was selected for the experiment. Other geometrical and operational constraints further impacted the experiment and they are presented hereafter:

Sky occultation by Earth: It is inferred from Figure 3.2 that none of the stars can be observed permanently since the Earth occulting zone will sweep over most of the sky in one orbit (the whole sky minus the two 30° half angle cones directed to the Sun and to the opposite). Here, the occulted zone represented by the blue ellipse will move up by 90° when the satellite spans the next quarter of orbit. A first option to go around this limitation could consist in pointing permanently within the sector around the Sun opposite direction: however this would violate not only the navigation requirement but also the power constraint since the solar panels would be almost perpendicular to the Sun. Another less radical option from the power point of view could consist in avoiding the Earth when its gets in the field of view by switching temporarily to a target located in the occultation-free sector and coming back to the favourable observation zone later on.

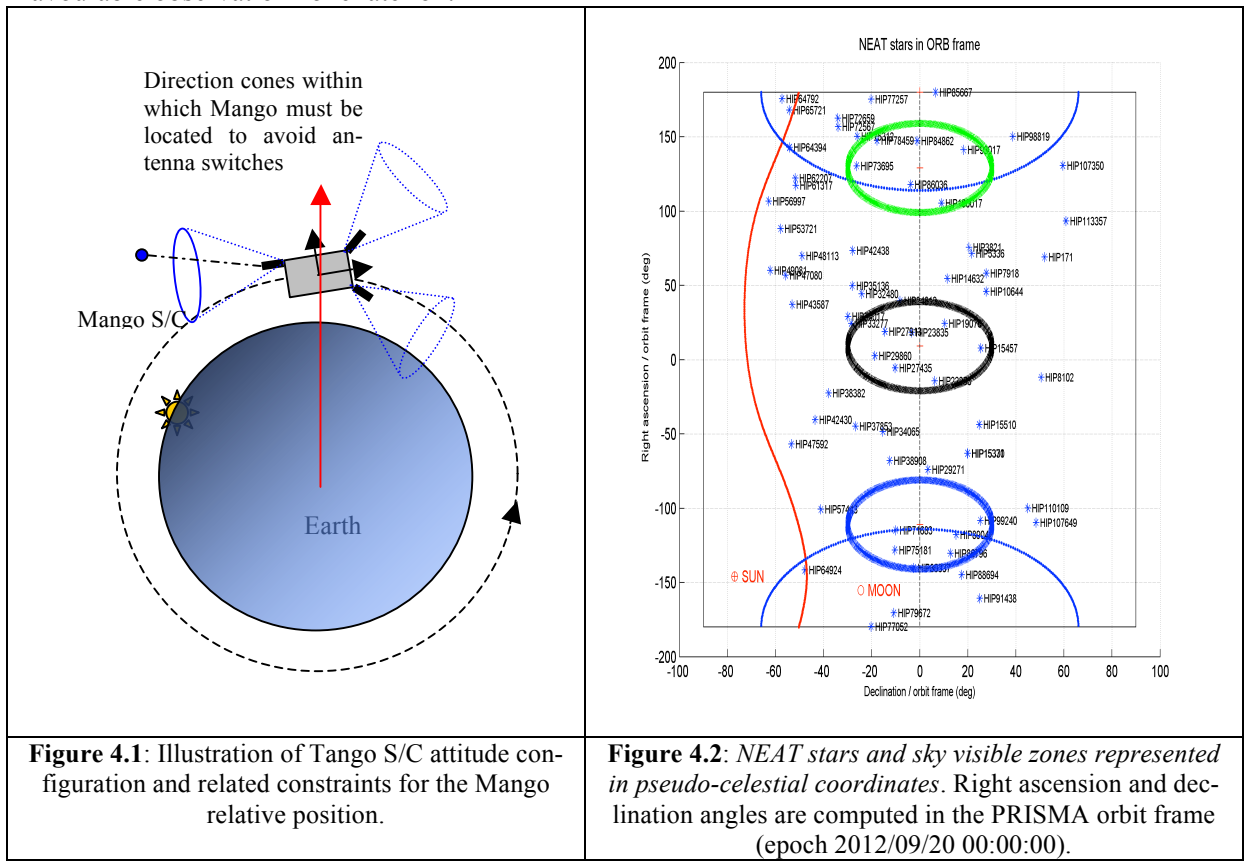

**Figure 4.1**: Illustration of Tango S/C attitude configuration and related constraints for the Mango relative position.

**Figure 4.2**: *NEAT stars and sky visible zones represented in pseudo-celestial coordinates*. Right ascension and declination angles are computed in the PRISMA orbit frame (epoch 2012/09/20 00:00:00).

This option was also discarded since this would imply large amplitude manoeuvres and would likely degrade the navigation performances. Finally, it was decided to live with this constraint specific to low earth orbits and to maintain the pointing during the occultation periods while guaranteeing for each target some minimum observation time.

Star tracker blinding: Mango S/C accommodates two star trackers (presented on Figure 1.a) that must not be both blinded for more than 25 consecutive minutes in order to avoid a transition in safe mode. In this experiment, such a blinding condition is due to occur when Earth gets in the back of Mango S/C. Given a list of celestial targets that satisfy the previous requirements, the only available leverages to maximize star tracker availability resides in the sizing of the observation / manoeuvre durations and in the sorting of the manoeuvres. Figure 6 shows the sequence of manoeuvres that was actually designed to observe the set of NEAT targets over a period of 3 orbits during the first experiment session. This sequence that allows each target to be observed and guarantees the best star tracker availability was obtained considering a fixed observation and manoeuvre duration for each target. These characteristics will be detailed in the sequel.

**Guidance & Control aspects**

During previous PRISMA experiments, position guidance and control was designed under the assumption that all motions were defined with respect to a local orbital frame generally attached to Tango S/C centre of mass. Forced trajectories were typically defined as combinations of station-keeping points and linear segments whereas control design relied on a model of the relative dynamic motion (Hill equations) and the use of LQR technique with feed-forward terms. This functionality package was not adequate to address the new manoeuvering needs. The GNC software had therefore to be adapted to allow the specification of motions in the inertial frame, the production of the appropriate guidance profile and the potential improvement of the control performance. However, this evolution had to be minimal to be compatible with the experiment limited ressources and it was decided to keep the same guidance and control paradigm by always converting the desired inertial motions into their counterparts in the local orbital frame.

Under this constraint, a fixed position in the inertial frame corresponds to a circular trajectory in the local orbital frame as it is illustrated on Figure 5.1. Since the desired location can be anywhere with respect to the orbital plane, the corresponding trajectory will not be coincident with the orbital plane but parallel with it. It will be defined by two parameters: the radius of the circle *R* and the cross-track distance *h.* These parameters can be expressed by the following formulas:

$$R = d.\left(\sqrt{u_X^2 + u_Z^2}\right) \quad (1)$$
$$h = d.u_Y \quad (2)$$

where $u = (u_X, u_Y, u_Z)^t$ represents the target direction vector projected in the local orbital frame at any given date and d is the inter-satellite distance.

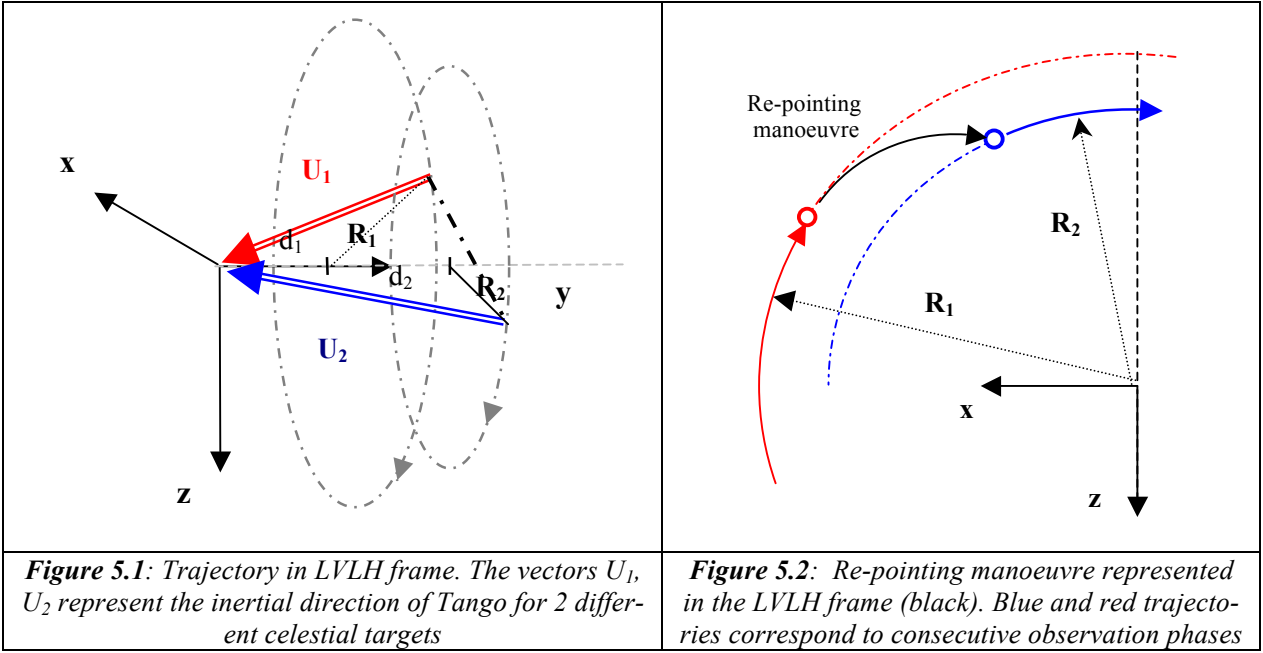

*Figure 5.1*: Trajectory in LVLH frame. The vectors $U_1$, $U_2$ represent the inertial direction of Tango for 2 different celestial targets

*Figure 5.2*: Re-pointing manoeuvre represented in the LVLH frame (black). Blue and red trajectories correspond to consecutive observation phases

*Guidance*: The guidance module accepts a trajectory input defined as a combination of linear segments and station keeping points with associated dates and produces a 6D position / velocity setpoint at 1 Hz cycle. Here, this module is adapted to extend the trajectory specification capability from the local orbital frame to the inertial frame. When inertial mode is selected, trajectory inputs are interpreted as inertial coordinates and the setpoints are actually computed in the inertial frame before being converted to the local orbital frame and delivered to the control module.

*Control*: Relative position control is based on a discrete LQR algorithm using a Linear Time Invariant relative dynamic model (Clohessy-Wilshire equations). The control structure includes also feed forward terms to compensate for quasi constant perturbations such as differential drag and solar radiation pressure. In addition, the LQR algorithm is designed to follow precisely a trajectory profile composed of constant velocity sections. The control structure is expressed as follows:

$$u = -[K_p . K_v]X + [M_p \ M_v].X_d + G \qquad (3)$$

where K, M and G represent respectively the 3x6 regulator gain matrix, the 3x6 input matrix and the feed forward terms. The typical period of activation has been so far either 100 s or 200 s. In the current experiment, the trajectory profile converted in the local orbital frame becomes a combination of circular arcs and portions of helices that are associated respectively to the observation periods and the re-pointing phases. Therefore, it is not possible anymore to keep the same control law as this was the case previously. However, the same control structure is compatible with the design of a specific control law for each type of trajectory element. Of course, this would imply to carry two sets of gains and swap them when entering or exiting each observation phase. For the sake of program simplicity, it was decided to keep a single set of gains that was adapted to optimize control performance during the observation phase.

*Control accuracy and propellant usage*: Assuming perfect navigation, control accuracy is driven by two main variables: the control rate and the thrust Minimum Impulse Bit (MIB). Trying to improve performance by reducing the control cycle is not adequate when the correction magnitude is likely to be smaller than the minimum impulse. Mango S/C MIB was 0.7 mm/s at the beginning of the mission (100 ms minimum pulse duration for a 1 N thruster and a 140 kg satellite mass). Fortunately, the thrust is smaller at the time of experiment due to the tank pressure reduction and this allows to select a control period of 100 s.

When a satellite is driven along a circular trajectory in the local orbital frame that is defined by the parameters R and d, the accelerations to be countered can be obtained by introducing the expression of the circular motion in the Clohessy-Wilshire equations:

$$A_X = -R.\omega.\cos \omega t$$
$$A_Y = d\omega^2 \qquad (4)$$
$$A_Z = -2R.\omega^2.\sin \omega t$$

An estimate of the delta-V budget required to maintain Mango S/C on a circular trajectory in the LVLH can be obtained by assuming a continuous thrust. The computation consists in integrating these perturbations over the desired time horizon. As an example, the theoretical delta-V budget to maintain Mango S/C on a 12 m radius circle with no cross-track bias is respectively 4.8 cm/s and 9.6 cm/s per orbit on the X and Z axes. However, we must take into account some loss of efficiency since the thrusters are not aligned with the thrust desired direction. The expected budget is therefore in the 16-17 cm/s range.

*Software evolution*: The CNES GNC algorithm had to be modified to perform the experiment and a software patch was uploaded just the day before. He software was produced starting from Matlab/Simulink models and using code generation tools (Mathwork Real Time Workshop). This approach allowed to prepare the experiment in a very short timeframe: the decision was taken end of July 2012, the algorithm was modified then validated with CNES and OHB-S simulation tools in August, software upload occurred in September 19$^{th}$.

**Experiment sessions**

The experiment is decomposed in two consecutive sessions performed on September 20th and 21st 2012. Each session starts from some station-keeping location in the vicinity of the companion spacecraft (Tango S/C) after a transfer from some stable relative orbit achieved under OHB-S control. The scenario actually begins when control is handed over to CNES and radio frequency navigation is initialized. Next, some forced translation along VBAR allows to reduce the distance to 14 m in about 2000s. Then, control is switched into inertial mode $t_0 + 2050$ s and the sequence of formation inertial manoeuvres is immediately initiated. The duration allocated to formation inertial pointing is purposely limited to 3 orbits for fuel saving considerations. The sequence is therefore completed five hours later ($t_0+ 20050$ s) and control is switched back into LVLH mode. At the end of the scenario (a few minutes later), control is handed over to OHB-S and Mango S/C is driven back to some paerking orbit.

The first session sticks as much as possible to the NEAT/µ-NEAT mission concept and involves real celestial targets from the NEAT catalogue – it relies also on FFRF navigation that is the candidate metrology for outer space missions. The scenario includes a sequence of 9 target time slots representative of a typical mission day except for the observation phase that is obviously downsized. Each target time slot lasts 2000 s including a 1400 s observation phase and a 600 s repointing manoeuvre

The second session involves a different and shorter set of targets (4) that includes this time the Moon for illustration purposes since the magnitude of NEAT targets is too faint for a direct observation with the cameras when the companion satellite is in the cameras field of view. The time slot is now 3000 s for the NEAT targets (2200 s for observation and 800 s for the repointing manoeuvre) whereas it is extended to 9000 s for the Moon observation. Control relies this time on relative GPS navigation to relax the experiment constraints and to offer some comparison benefits. In both sessions, the manoeuvre sequence has been optimized to limit the star tracker masking by the Earth and its impact on the attitude estimation performance.

*Table 1 : NEAT stars observed during the first rehearsal*

| Basic information (Hipparcos catalogue 1997) | | | | | |
|---|---|---|---|---|---|
| Star_identifier | HD ident | RA (J 2000) | DEC (J 2000) | Vmag | SpType |
| HIP 15457 | 20 630 | 03 19 21.696 | +03 22 12.71 | 4,84 | G5Vvar |
| HIP 19076 | 25 680 | 04 05 20.258 | +22 00 32.06 | 5,90 | G5V |
| HIP 22263 | 30 495 | 04 47 36.292 | -16 56 04.04 | 5,49 | G3V |
| HIP 23835 | 32 923 | 05 07 27.006 | +18 38 42.19 | 4,91 | G4V |
| HIP 24813 | 34 411 | 05 19 08.474 | +40 05 56.59 | 4,69 | G0V |
| HIP 27435 | 38 858 | 05 48 34.941 | -04 05 40.73 | 5,97 | G4V |
| HIP 27913 | 39 587 | 05 54 22.982 | +20 16 34.23 | 4,39 | G0V |
| HIP 29860 | 43 587 | 06 17 16.138 | +05 06 00.40 | 5,70 | G0.5Vb |
| HIP 33277 | 50 692 | 06 55 18.668 | +25 22 32.51 | 5,74 | G0V |
| HIP 34017 | 52 711 | 07 03 30.459 | +29 20 13.49 | 5,93 | G4V |

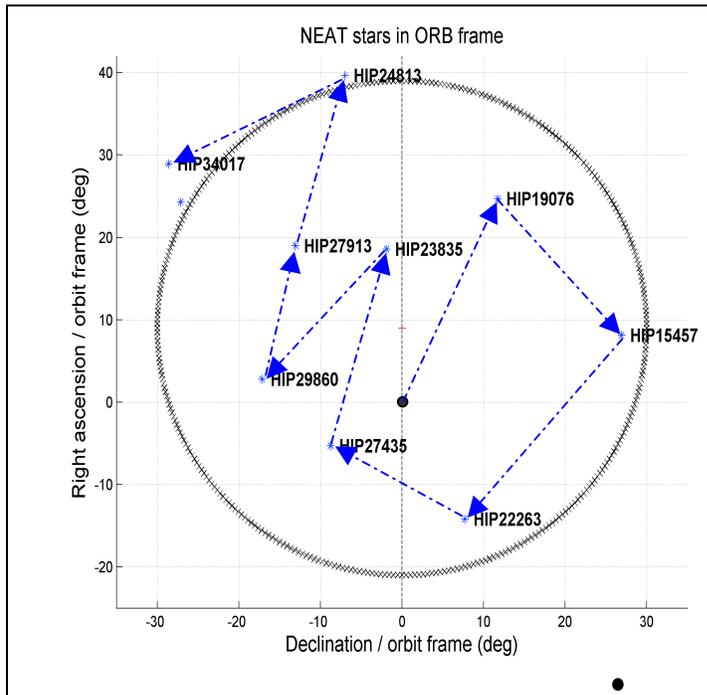

| Man Id | Manœuvre magnitude (°) | Translation magnitude (m) |
|---|---|---|
| 1 | 26.8377 | 5.41 |
| 2 | 21.7218 | 4.44 |
| 3 | 28.5856 | 5.74 |
| 4 | 18.7034 | 3.84 |
| 5 | 24.7205 | 5.02 |
| 6 | 21.7623 | 4.45 |
| 7 | 16.1635 | 3.34 |
| 8 | 21.2071 | 4.34 |
| 9 | 23.8714 | 4.85 |

**Figure 6.1** : *Sequence of manoeuvres to maximize star tracker availability. The map origin corresponds to the zenith direction at the orbit ascending node. Vertical axis coincides with the orbital plane.*

**Figure 6.2**: *Manoeuvre characteristics associated to the sequence in Figure 5.1.*

# FLIGHT RESULTS

*Results overview*: Both experiment sessions have been monitored carefully during the various passages and no anomaly prevented them to run until completion. Fuel usage was particularly scrutinized and it always remained in line with expectations (about 30 cm/s per orbit). Control behaviour was checked using the alternate navigation function that ran in open loop (based on either GPS or FFRF measurements) and did not raise any particular worry. Star tracker availability stayed on the safe side with some satisfactory margins. Images acquired periodically allowed to get a direct confirmation of the perfect functional behavior. The only concern came from one thruster temperature that exceeded the operational limit and the experiment interruption was envisioned. The risk was deemed acceptable since the manufacturer quickly confirmed the existence of some confortable margin and also because the thruster that was heavily sollicited was due to become idle soon after the next re-pointing manoeuvre.

Control performance has been being assessed using three navigation references: RF navigation, on board GPS navigation and Precise Orbit Determination (POD). POD is generally the most accurate reference when the pointing of both satellites is quasi constant in a local orbital frame and a 1 cm (1 $\sigma$) 3D accuracy is achievable. Conversely, the navigation conditions get far from optimal in inertial pointing and some performance degradation can be expected. Such a situation is actually observed in both experiments since POD cannot be properly reconstructed for about 50% of the inertial pointing period.
Control performance presented on Table 2 concern only the observation phases: the time slots associated to the repointing manoeuvres are ignored when computing the error signal statistics (mean and standard deviation parameters).

**Table 2: Summary of experiment performances**

| Experiment | Delta-V (cm/s) | Control error (m) RF nav as ref. | Control error (m) GPs nav as ref. | Control error (m) POD as ref. (*) |
|---|---|---|---|---|
| Session 1 | 95.4 | **Bias [1.13 5.03 1.83]** **Std [3.78 2.42 3.31]** | Bias [7.32 12.98 6.94] Std [8.07 23.4 18.27] | Bias [9.78 8.46 6.31] Std [5.43 6.76 5.42] |
| Session 2 | 65.2 | N/A | **Bias [1.09 4.09 1.39]** **Std [4.48 7.51 4.78]** | Bias [-9.35 8.03 -5.87] Std [5.87 15.83 11.57] |

Bold figures correspond to performances that are observed with the navigation function used in the control loop. These figures illustrate the control behavior but do not reflect the actual positioning performances that take into account the navigation error budget. These latter performances are usually provided by POD but this data when available cannot be blindly trusted given the poor GPS measurement conditions.

*1st session*: This session that relies on RF navigation allows to obtain the best control performances during the observation periods. These are actually consistent with the ones obtained during previous station-keeping experiments performed in LVLH frame where the control stability was in the 2 cm range [7]. The relatively good behaviour despite higher gravity gradient perturbations can be explained by the MIB reduction.

Some mismatch between RF and GPS navigation can be observed on Figure 7.2. A 1.5 m spike is visible around 16:30 and can be attributed to bad GPS measurements (the number of common GPS satellites is reduced to 3 for a few minutes). POD manages to handle the situation and confirms the correct RF navigation behaviour. Figure 6.2 shows also the sub optimal control behaviour during the re-pointing manoeuvres with control errors increasing up to 20 cm. However, this large error is quickly reduced during the first control cycle of the observation phase. As regards the total propellant budget, it looks rather high for 3 orbits (almost 1 cm/s) but a large part of it comes actually from the large amplitude re-pointing manoeuvres. Considering the observation phases only, this budget goes down to roughly 18 cm/s per orbit. On the other hand, the experiment is a bit disappointing from the illustration point of view since the selected targets sere too faint to be observed on the FR VBS and DVS cameras.

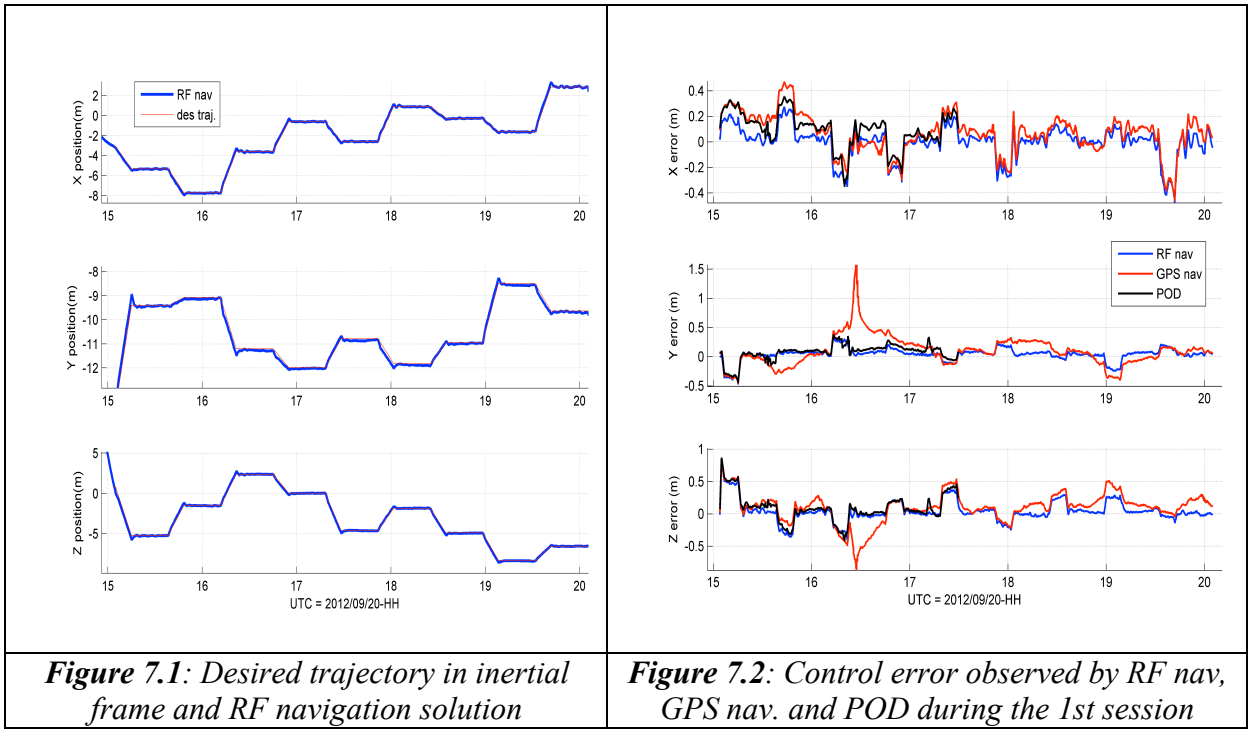

*Figure 7.1*: Desired trajectory in inertial frame and RF navigation solution   *Figure 7.2*: Control error observed by RF nav, GPS nav. and POD during the 1st session

*Second session*: This session that relies on GPS navigation benefits from better conditions since both re-pointing manoeuvres and observation periods are longer. However, control performances are coarser due to some GPS navigation degraded behaviour. Figure 8.2 illustrates the difference between on board GPS navigation and POD that can get in the 30 cm range with sudden variations that are due to affect negatively the control algorithm.

This effect is not obvious on the propellant budget since it is 65 cm/s only or 30 cm/s less than the first session. This saving is explained by the smaller magnitude of the re-pointing manoeuvres and their longer duration. The budget associated to the observation phases is actually identical to the first session: an average 18 cm/s per orbit.

RF navigation data that is not available in this mode could not be reconstructed *a posteriori* due to missing telemetry and navigation comparison was unfortunately not possible. Here, the design was driven by the illustration objective (the moon is chosen as target) and not the navigation / control performance. RF navigation was a better choice to achieve performance but the risk was higher given the direction of the targets considered.

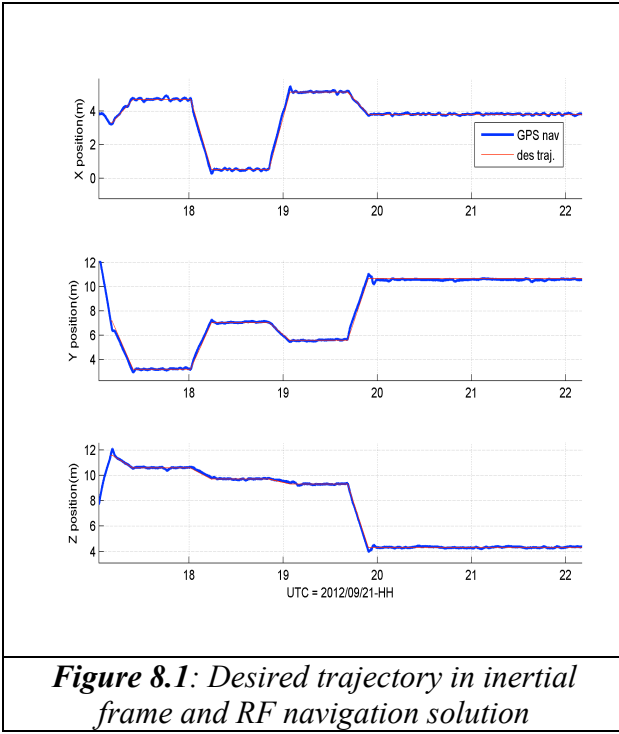

*Figure 8.1*: *Desired trajectory in inertial frame and RF navigation solution*

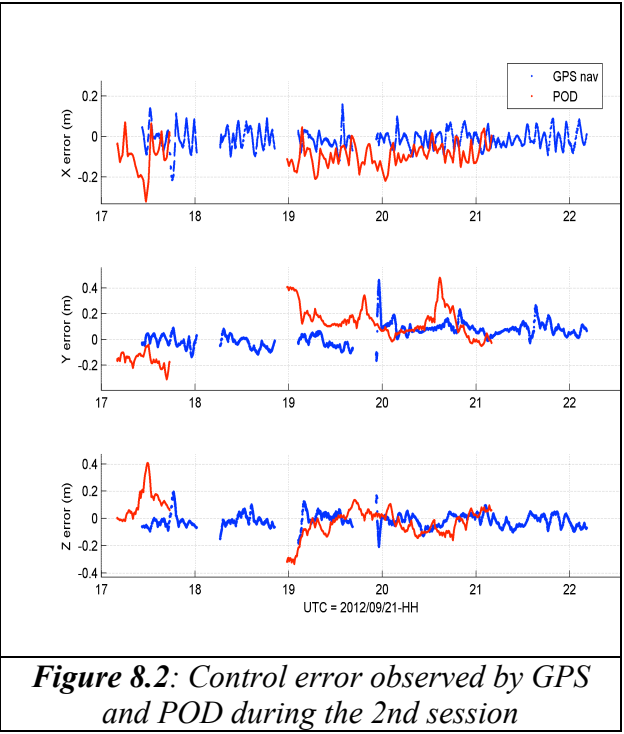

*Figure 8.2*: *Control error observed by GPS and POD during the 2nd session*

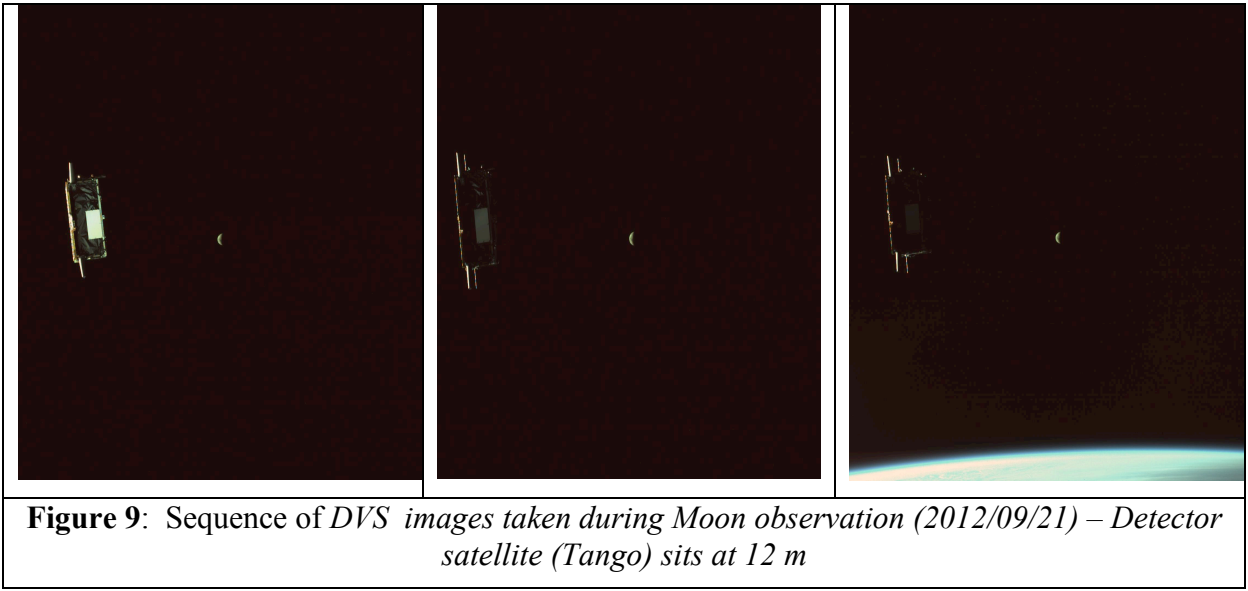

**Figure 9**: Sequence of *DVS images taken during Moon observation (2012/09/21) – Detector satellite (Tango) sits at 12 m*

**CONCLUSION**

This paper has described the first demonstration of formation flying in inertial frame that was ever performed. This experiment that was achieved successfully consisted in mimicking the type of manoeuvres that will be required in future astrometry missions represented by the NEAT/µ-NEAT con-cepts. The bi-satellite formation was pointed successively to a set of celestial targets and the configuration was maintained fixed during a period of time representing the observation phase.

The functional behaviour has been satisfactory and allowed to reach the expected level of positioning performances. Control stability better than 4 cm (1σ) could be obtained during the observation periods and this represents a significant achievement considering the stringent environmental and operational constraints. The experiment objective was to push the whole flight system to its limits and it actually got close to the breaking point. Some performance improvement was still within reach but would have required some additional time for scenario adjustment. In particular, the second session offered better conditions to achieve accurate positioning due to longer observation phases but using GPS instead of RF navigation did not allow benefiting from it.

The main performance limitation comes obviously from the PRISMA platforms metrology and actuation devices. First of all, Tango S/C attitude variations degrade RF measurements accuracy by a modification of the bias due to multipath. Then, the thruster minimum impulse is not small enough to correct anytime the observed position offsets. Considering the results obtained from the different PRISMA experiments, RF metrology appears compatible with a 1 cm positioning accuracy in Low Earth Orbit assuming that the satellite companion is properly stabilized and fine actuation is available.

The good functional behaviour and the level of performance achieved despite the numerous constraints constitutes a blatant illustration of the margins that should be available when going on a higher orbit and using finer equipment for both metrology and actuation as it is envisioned for the future Proba3 technology mission. This experiment is definitely paving the way to such demonstrations: most of the functionalities have been already exercised; the next step will consist in bringing the actual positioning accuracy down to the millimetre range.

Even though formation flying was the first application target, it must be outlined that this experiment addressed also indirectly the field of orbital debris removal. Turning around space debris to perform inspection and potentially cancel some angular rate before capture will definitely rely on different metrologies. Nevertheless, the same guidance and control functionalities will be required and scenario design will have to cope with similar constraints such as star tracker blinding, possible thruster overheating and power shortage. Finally, this flawless experiment that was developed in a very short time frame has shown again the great potential and flexibility of the PRISMA test bed environment.

## 7. Acknowledgements

Our acknowledgements must be conveyed to CNES for financing this activity and OHB-Sweden for its outstanding technical support during the experiment preparation as well as the flight operations.